\documentclass[letter]{aa} % for the letters
\usepackage{graphicx}
\usepackage{txfonts}
\usepackage{enumitem}
\usepackage{amsmath,amssymb}
\usepackage{booktabs}
\usepackage{xspace}
\usepackage{textcomp}
\usepackage{xcolor}	
\usepackage[euler]{textgreek}
\usepackage{dirtytalk}
\usepackage{caption}
\usepackage{subcaption}
\graphicspath{{./figures}}
\newcommand{\oiii}{[\ion{O}{III}]\,88\,\textmu m}

\newcommand{\ofivetwo}{[\ion{O}{III}]\,52\,\textmu m}

\begin{document}

   \title{An upper limit on \oiii\ and 1.2~mm continuum emission from a JWST $z\approx12-13$ galaxy candidate with ALMA}

   \titlerunning{An upper limit on \oiii\ and 1.2~mm continuum emission from a JWST $z>12$ galaxy candidate}
   \author{Gerg\"o Popping
          \inst{1}
          }

   \institute{European Southern Observatory, Karl-Schwarzschild-Str. 2, D-85748, Garching, Germany\\
              \email{gpopping@eso.org}
         % \and
         %     University of Alexandria, Department of Geography, ...\\
         %     \email{c.ptolemy@hipparch.uheaven.space}
         %     \thanks{The university of heaven temporarily does not
         %             accept e-mails}
             }

   %\date{Received ...; accepted ...}

% \abstract{}{}{}{}{}
% 5 {} token are mandatory
 
  \abstract
   {A number of new $z>11$ galaxy candidates have recently been identified based on public James Webb Space Telescope (JWST) NIRCam observations. Spectroscopic confirmation of these candidates is necessary to robustly measure their redshift and put them in the context of our understanding of the buildup of galaxies in the early Universe. GLASS-z13 is one of these candidates, with a reported photometric redshift $z>11.9$. %, a stellar mass of $\log{(M_{\rm star}/\rm{M}_\odot)} = 9.0^{+0.3}_{-0.4}$ and star-formation rate (SFR) averaged over the last 50 Myr of $7^{+4}_{-3}\,\rm{M}_\odot\,\rm{yr}^{-1}$. 
   I present publicly available Atacama Large Millimeter/submillimeter Array (ALMA) band 6 Director's Discretionary Time observations (project 2021.A.00020.S; PI T. Bakx), taken to acquire a spectroscopic redshift for GLASS-z13 by searching for \oiii\ line emission in the redshift range $z=11.9-13.5$. No \oiii\ emission is detected in integrated spectra extracted within an aperture around GLASS-z13, nor when using an automated line finding algorithm (applying different uv-weighting strategies for the imaging). 1.2~mm continuum emission associated to GLASS-z13 is not detected either. If GLASS-z13 is at z$\approx$12-13, this implies a 3-$\sigma$ upper limit on the  \oiii\  and rest-frame $\sim$90 $\mu$m continuum emission of $\sim1\times10^8\,\rm{L}_\odot$ and 10.8 $\mu$Jy, respectively. The non-detection of \oiii\ and continuum emission does not necessarily imply that GLASS-z13 is not at $z\approx12-13$. It can also be explained by a low metallicity  ($\sim 0.2\,\rm{Z}_\odot$ or lower) and/or high-density (at least 100 $\rm{cm}^{-3}$) interstellar medium. This work demonstrates the synergy between ALMA and JWST to study the properties of the first galaxies, although JWST/NIRSpec spectroscopy will  be necessary to confirm or reject the high photometric-redshift of GLASS-z13. 
  }


   \maketitle
%
%________________________________________________________________

\section{Introduction}
One of the main science goals of the James Webb Space Telescope (JWST) is  detecting the very first galaxies that formed after the Big Bang. Making use of the public release of the first JWST observations, a number of papers have reported new $z>10$ galaxy candidates based on spectral energy distribution (SED) fitting of JWST near-infrared photometry \citep{Adams2022,Atek2022,Bouwens2022,Bradley2022,Castellano2022,Donnan2022,Harikane2022b,Hsiao2022, Finkelstein2022,Labbe2022,Naidu2022,Naidu2022b,Yan2022, Finkelstein2022b,Rodighiero2023}. 

The highest-redshift spectroscopically confirmed galaxy prior to the launch of JWST was GN-z11 at $z=10.957$ \citep{Oesch2016,Jiang2021}, with a reported UV luminosity of $M_{UV} = -22.1\pm0.2$. Recently, \citet{Harikane2022} presented a tentative 3.8$\sigma$ detection of the \oiii\ emission line of a $z=13.27$ galaxy candidate with a UV luminosity of $M_{UV} = -23.3$ (called HD1; though \citet{Kaasinen2022} argue through statistical tests that this detection is fully consistent with being a noise fluctuation). The number density of these two bright  $z>10$ (candidate) galaxies is already higher than expected from theoretical models \citep[e.g.,][]{Yung2019, Behroozi2020}. A spectroscopic confirmation of the high redshift of the newly reported bright JWST $z>10$ candidates is thus crucial to  provide further robust constraints for galaxy formation models and further our understanding of the buildup of the very first galaxies in our Universe  \citep{Ferrara2022, Inayoshi2022, Liu2022, Lovell2022,Mason2022, MKB, Mirocha2022}. \citet{CurtisLake2022} recently presented spectroscopic redshift confirmation of four galaxies at $z=10.45$ to 13.2, with UV magnitudes in the range between -18.2 and -19.3, based on JWST/NIRSpec spectroscopy.

If all reported JWST high-redshift UV-bright candidates are indeed at $z>10$ this would imply little to no evolution in the UV luminosity function of galaxies from $z=8$ to $z=12$ (e.g., \citealt{Naidu2022}, see also \citealt{Bowler2020}) and highly efficient star formation within the first few 100 Myrs after the Big Bang. Spectroscopic confirmation of their redshift is necessary, since high redshift candidates classified based on near-infrared (NIR) photometry alone may also be dusty star-forming galaxies (DSFG, e.g., \citealt{Zavala2022}) or quenched galaxies at $z=3-6$ (see the alternate solution for HD1 in \citealt{Harikane2022}). 

\begin{figure*}
	\centering
	    \includegraphics[width=\linewidth]{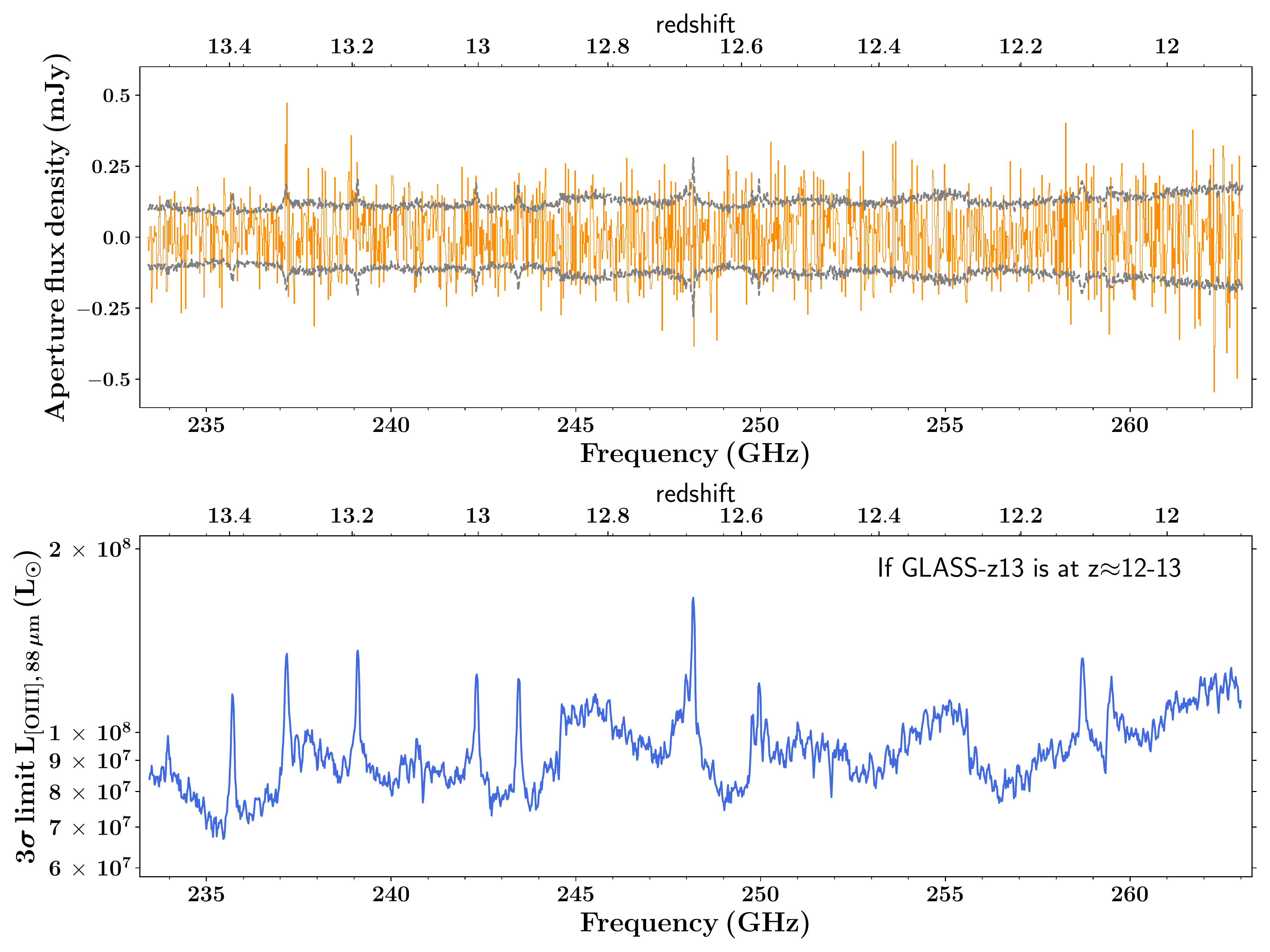}
     \caption{\label{fig:spectral_scan}Top row: The ALMA spectrum of GLASS-z13. The spectrum was created by extracting the flux density within a circular aperture with a diameter of 0.5 arcsec around GLASS-z13. The orange bars indicate the spectrum, whereas the grey dashed lines correspond to the one-$\sigma$ noise limit of the data. No emission line is detected within the spectrum. Bottom row: The three--$\sigma$ limit on the \oiii\ luminosity of GLASS-z13 as a function of redshift  if GLASS-z13 is actually at $z\approx12-13$, but its \oiii\ luminosity is too faint to have been detected by ALMA. The upper limit on the \oiii\ emission is calculated assuming a line-width of 100 km s$^{-1}$, by taking the integrated noise within a window of width 100 km s$^{-1}$ around the frequency of interest.}
\end{figure*}

One particularly exciting $z>10$ candidate was recently reported by \citet{Castellano2022} and \citet{Naidu2022}. This galaxy was observed as part of the GLASS-JWST Early Release Science Program \citep{Treu2022} and is dubbed GLASS-z13 in this paper. Independent spectral energy distribution (SED) fitting of the Lyman-alpha break of this source based on JWST/NIRCam photometry from approximately 1 to 5 $\mu$m by \citet{Castellano2022} and \citet{Naidu2022} places this galaxy at $z\approx12-13$, with a UV luminosity of $M_{UV} = -21.0\pm.1$. %This high redshift  was obtained using three different SED fitting codes, namely \texttt{Prospector} \citep{Leja2017,Leja2019,Johnson2021}, \texttt{zphot} \citep{Fontana2000} and \texttt{EaZy} \citep{Brammer2008}. 
\citet{Naidu2022} and \citet{Santini2022} perform SED fitting to the available photometry for GLASS-z13 and find a stellar mass of $\log{(M_{\rm star}/\rm{M}_\odot)} = 9.1^{+0.3}_{-0.4}$ and $8.1^{+0.4}_{-0.1}$ and SFR of $6^{+5}_{-2}$ and $19.1^{+13.5}_{-10.1}\,\rm{M}_\odot\,\rm{yr}^{-1}$, respectively. These differences may be caused by the use of different SED fitting codes. A further relevant systematic uncertainty on the derived properties, including the photometric redshift of GLASS-z13, is the uncertainty in the zero-point of the NIRCam flux-calibration (e.g., see Section 2.2 in \citealt{Adams2022}. Spectroscopic confirmation is thus necessary to overcome the uncertainties on the photometric redshift. %Spectroscopic confirmation is nevertheless necessary to rule out a scenario where GLASS-z13 is a lower redshift DSFG or quenched galaxy. 
%If its high redshift is spectroscopically confirmed, the existence of this UV-bright galaxy at $z\approx12-13$ strengthens the hypothesis that there is little to no evolution in the bright end of the UV luminosity function from z$=8$ to z$\sim12$ \citep{Naidu2022}. 

In this paper I present publicly available Atacama Large Millimeter/submillimeter Array (ALMA) band 6 data taken as part of a recently approved Director's Discretionary Time (DDT) program (2021.A.00020.S; PI: T. Bakx). The observations were targeted towards GLASS-z13 with the aim of detecting the \oiii\ emission line to obtain a robust spectroscopic redshift for GLASS-z13. This strategy has successfully been adopted in the last years to obtain spectroscopic redshifts of $z\sim9$ galaxies \citep[e.g.,][]{Laporte2017,Hashimoto2018, Tamura2019,Laporte2021} and recently been used to search for \oiii\ emission from various other JWST selected $z>10$ galaxy candidates \citep{Yoon2022,Fujimoto2022}. A similar analysis on GLASS-z13 is furthermore presented in \citet{Bakx2022}. In Section~\ref{sec:data} I present the ALMA data. I present the data analysis in Section~\ref{sec:analysis} and discuss the findings in Section~\ref{sec:discussion}. A summary of the results is presented in Section~\ref{sec:summary}. Throughout this paper I use a flat $\Lambda$-CDM concordance model (H$_0 =$ 70.0 km s$^{-1}$ Mpc$^{-1}$, $\Omega_M =$ 0.30).

\begin{figure}
	    \includegraphics[width=\linewidth]{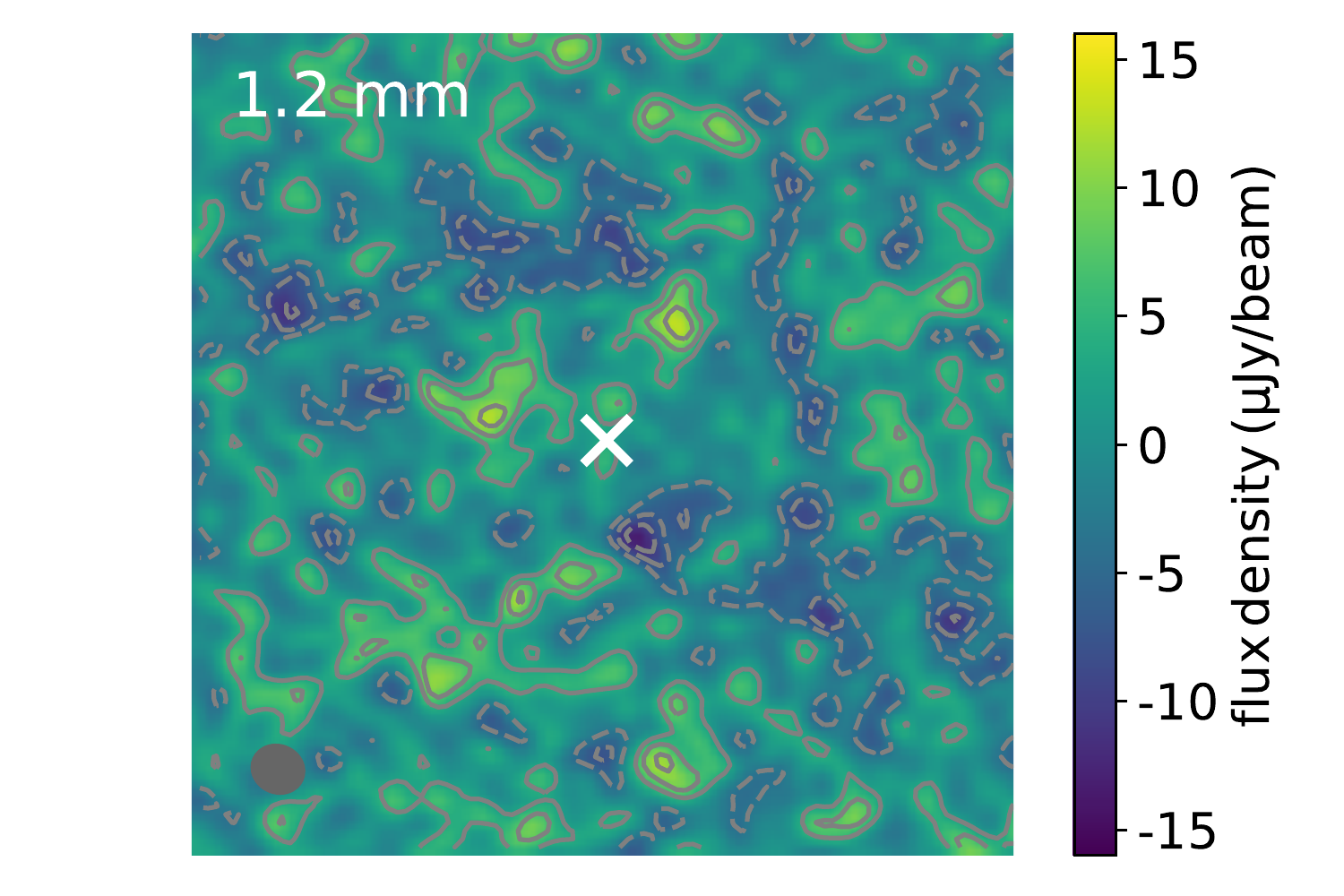}
     \caption{\label{fig:poststamp}A 5" $\times$ 5" cutout of the ALMA 1.2~mm continuum map around GLASS-z13. Contours indicate the -3, -2, -1, 1, 2 and 3 $\sigma$ levels. The beam of the ALMA 1.2~mm image is indicated with the grey ellipse in the bottom left corner. The location of GLASS-z13 is indicated with a white cross. No continuum emission is detected at the location of GLASS-z13. }
\end{figure}

\section{ALMA Observations and Data Reduction}
\label{sec:data}
In this paper I use publicly available ALMA band 6 data obtained as part of project 2021.A.00020.S (PI: T. Bakx) on August 3, 4 and 5, 2022 in configuration C-5. The water vapour during the observations was in the range PWV = 0.5 - 1 mm. The data cover a contiguous frequency range from 233.41 GHz to 263.05 GHz with a spectral resolution of 7812.01 kHz. This frequency coverage corresponds to \oiii\ emission at redshifts from $z=11.9$ to $z=13.5$. 

I make use of calibrated measurement sets restored using the Common Astronomy Software Applications package \citep[CASA,][version 6.2.1]{McMullin2007} calibration scripts provided by the ALMA observatory. A spectral cube covering the entire frequency range of the observations was generated with the \texttt{tclean} task with a spectral resolution of 20 MHz (corresponding to $\sim$ 25 km s$^{-1}$ for a typical rest-frequency of 250 GHz) and applying Briggs weighting with robust equals 2 (natural weighting). The resulting cube has a beam size of  0.39 arcsec $\times$ 0.34 arcsec and a typical noise level of $0.15\,\rm{mJy}/\,20\,\rm{MHz}/\,\rm{beam}$. The same weighting parameters were used to create a 1.2 mm band 6 continuum image of GLASS-z13, resulting in a beam size of 0.34 arcsec $\times$ 0.31 arcsec and a typical noise level of 3.6 $\mu$Jy per beam. I furthermore present a uv-tapered cube  (with a taper of 0.6 arcsec) and image with a beam size of 0.8 arcsec to potentially identify low-surface brightness emission missed in the images  with a smaller beam. The results discussed in this paper are qualitatively the same between the images created with the various weighting choices.

\section{No \oiii\ and 1.2~mm continuum emission detection}
\label{sec:analysis} 
The aim of this paper is to search for \oiii\ emission associated to GLASS-z13. I obtain an ALMA flux density spectrum by extracting the flux density along the frequency axis within a circular aperture with a diameter of 0.5 arcsec around the expected position of GLASS-z13 based on the JWST/NIRCam photometry. This diameter corresponds to approximately 1.25 times the beam size of the observations and a physical scale larger than the NIR half light-radius derived for GLASS-z13 of 0.5 kpc \citep{Naidu2022}. The noise level for this aperture is estimated from the non-primary beam-corrected cube, by taking the standard deviation of the flux density in 100 apertures of the same size and offset randomly from the source position for every individual channel. The resulting integrated aperture flux density spectrum is presented in the top panel of Figure \ref{fig:spectral_scan}. No emission line is visible in this spectrum. This conclusion is robust against changes in the aperture diameter between 0.4 and 1.0 arcsec and re-binning of the data to 40 MHz per channel ($\sim 50\,\rm{km}\,\rm{s}^{-1}$). I do not find an emission line in the spectrum extracted from the uv-tapered cube either (see Figure~\ref{fig:spectral_scan_tapered} in Appendix~\ref{sec:appendix}). These results are consistent with the findings by \citet{Bakx2022}, who do not find \oiii\ line-emission at the location of GLASS-z13 either.

I use the \texttt{Findclump} algorithm \citep{Walter2016} that was designed to identify emission lines in an automated fashion to search for a line using a more objective approach. The algorithm was set up to search for lines in kernels of 3 up to 19 channels, cropping those sources with a signal-to-noise ratio larger than 3 that fall within 1 arcsec and 0.4 GHz  (corresponding to $\sim$500  km s$^{-1}$) of each other. The fidelity of the detected lines (probability of the line to be real) is estimated by comparing the number of positive lines to the number of negative lines with a signal-to-noise ratio of at least 3 (see \citealt{Walter2016} for a description). No reliable emission line is detected within 10 arcsec of the location of GLASS-z13. This is in agreement with the findings based on Figure \ref{fig:spectral_scan}. No emission line is detected in the uv-tapered cube either. The lack of detected \oiii\ emission suggests that no \oiii\ emission line is present in the explored frequency range, or it is fainter than the sensitivity limit of the data. 

\citet{Bakx2022} report a tentative $\sim 5\,\sigma$ extended detection approximately 0.5" offset from the location of GLASS-z13. \texttt{Findclump} identifies this line, but assigns it a fidelity of only 0.04 (probability of the line being real of 4\%).

In the bottom panel of Figure~\ref{fig:spectral_scan} I show the upper limit that can be derived on the \oiii\ luminosity, assuming that GLASS-z13 is at $z\approx12-13$ and that the \oiii\ emission line falls in the frequency range covered by the observations. I adopt a line width for the \oiii\ emission of 100 km s$^{-1}$, similar to \oiii\ line widths reported  for $z\sim8-9$ galaxies \citep{Hashimoto2018, Tamura2019}. For a channel width of 25 km s$^{-1}$ (dividing the \oiii\ emission over four channels) the  three $\sigma$ limit on the \oiii\ line luminosity is of the order $8\times10^7$ and $1 \times 10^8 \,\rm{L}_\odot$ at 240 and 260 GHz, respectively. If we instead adopt a larger line width of 400 km s$^{-1}$ (see for example \citealt{Hashimoto2019} and \citealt{Wong2022}) the three $\sigma$ upper limit increases by a factor of two, up to $2 \times 10^8 \,\rm{L}_\odot$  at 260 GHz.

%Recently, \citet{Zavala2022} used a 1.1 mm continuum detection of a $z>10$ galaxy candidate to reveal that this candidate may actually be a DSFG at $z\sim$5.
In Figure \ref{fig:poststamp} I present the ALMA 1.2~mm continuum image around GLASS-z13. No 1.2 mm continuum emission associated to GLASS-z13 is visible and hence a 3-$\sigma$ upper limit of 10.8 $\mu$Jy can be placed. I do not find 1.2~mm continuum emission associated to GLASS-z13 in the uv-tapered continuum map either (Figure~\ref{fig:poststampTapered} in Appendix~\ref{sec:appendix}).

\begin{figure*}
	    \includegraphics[width=\linewidth]{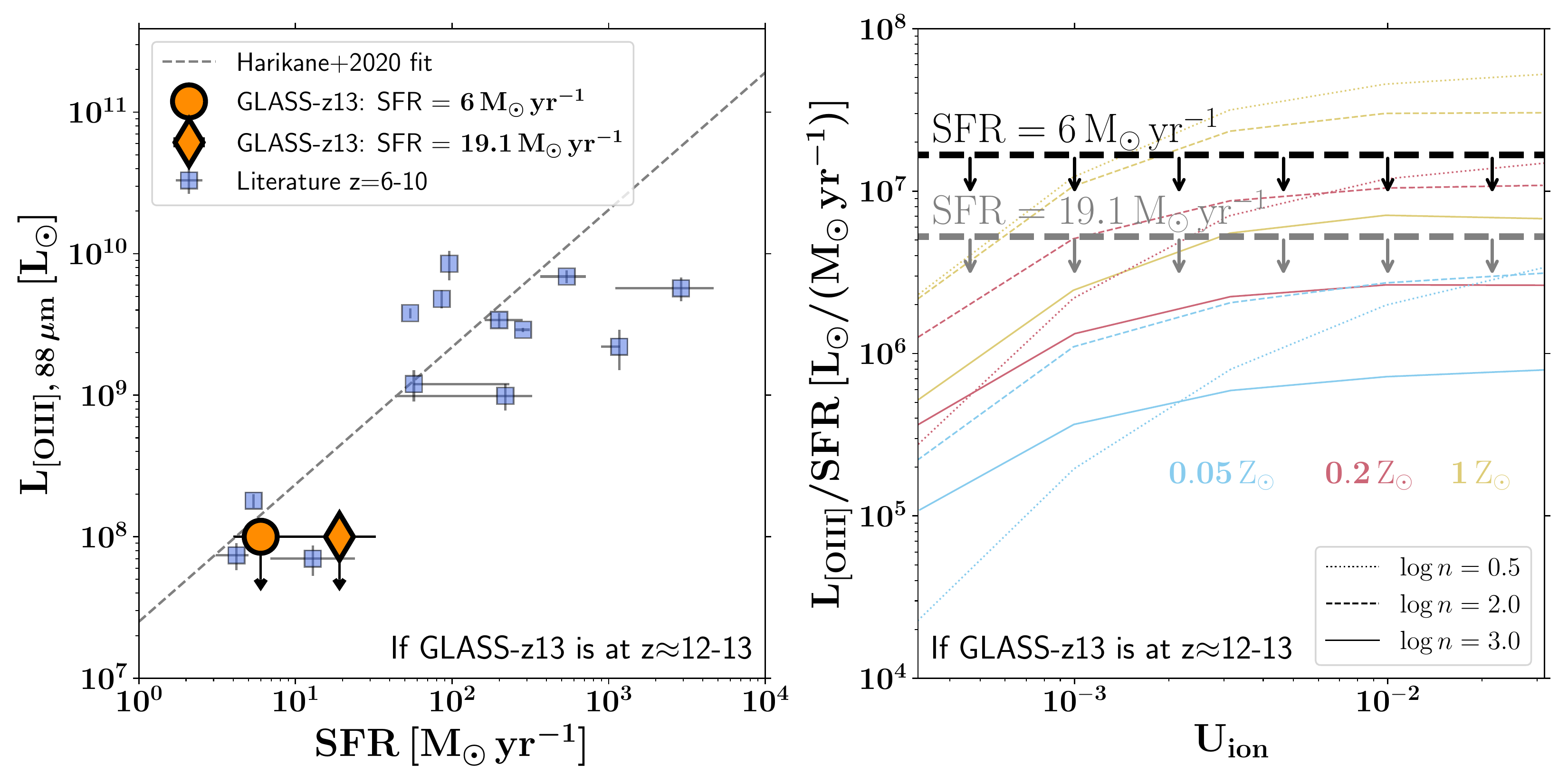}
     \caption{\label{fig:oiii-sfr} Left: Upper limit on the \oiii\ luminosity of GLASS-z13 as a function of its SFR, compared to galaxies at $z=6-10$ taken from the literature compilation by \citet{Harikane2020}. Upper limits are presented for the two respective SFRs derived by \citet{Castellano2022} and \citet{Santini2022}. If GLASS-z13 is a $z>11.9$ galaxy, the upper limits are consistent with the empirical trend. Right: CLOUDY calculation results taken from \citet{Harikane2020} for the ratio between \oiii\ luminosity and SFR, as a function of ionization parameter $\rm{U}_{\rm ion}$, ISM metallicity and density. Under the assumption that GLASS-z13 is a $z\approx12-13$ galaxy, the upper limits on the \oiii\ luminosity--SFR ratio that would be obtained for the two SFRs derived in the literature are indicated by the dashed black and grey lines with arrows. A broader line width of $\sim 400\,\rm{km}\,\rm{s}^{-1}$ would move the upper limits up by a factor of two. These upper limits suggest either a low-metallicity ($<0.2\,\rm{Z}_\odot$) and/or a dense (at least 100 cm$^{-3}$) ISM.}
\end{figure*}

\section{Discussion} % (fold)
\label{sec:discussion}
The non-detection of \oiii\ and 1.2~mm dust-continuum emission associated to GLASS-z13 can be explained by three scenarios that are discussed in the subsections below.

\subsection{GLASS-z13 is a $z\approx12-13$ galaxy with an \oiii\ luminosity below the sensitivity limit of the ALMA data}
The first scenario to explain the non-detection of \oiii\ emission is that the line does fall in the frequency range covered by the observations, but is too faint to have been detected. To put this scenario in context I compare in the left panel of Figure~\ref{fig:oiii-sfr} the 3-$\sigma$ \oiii\ limit for GLASS-z13 ($\sim6\times10^{6}\,\rm{L}_\odot$; see Section~\ref{sec:analysis}) to the \oiii\ luminosities measured for galaxies at $z=6-9$ as a function of their SFR (from the compilation presented in \citealt{Harikane2020}). The  upper limit on the \oiii\ luminosity for GLASS-z13 is consistent with the empirical relation between the \oiii\ luminosity and SFR of galaxies (regardless of the SFR taken for GLASS-z13).

\citet{Harikane2020} explored the ratio between the \oiii\ line-luminosity of galaxies and their SFR as a function of gas-phase metallicity, density and ionisation parameter $U_{\rm ion}$ using \texttt{CLOUDY} \citep{Ferland2017} modeling. In the right panel of Figure~\ref{fig:oiii-sfr} I compare the upper limit on the \oiii\ luminosity--SFR ratio to the model predictions taken from \citet{Harikane2020}. If GLASS-13 is at $z\approx12-13$, the upper limit on the \oiii\ luminosity--SFR ratio can be explained by low-metallicity gas ($\sim0.2\,\rm{Z}_\odot$ or lower). A metallicity of $0.2\,\rm{Z}_\odot$ or less is fully consistent with recent model predictions for the gas-phase metallicity of $z=12-13$ galaxies with a stellar mass of $\sim 10^9\,\rm{M}_\odot$ \citep{Wilkins2017,Ucci2021}. A dense ISM of  $10^3\,\rm{cm}^{-3}$ or possibly even higher provides an alternative explanation for the non-detection of \oiii\ emission (even in combination with a solar-metallicity ISM). Indeed, \citet{Jiang2021} derived a density for GN-z11 (a $z\sim 11$ spectroscopically confirmed galaxy) of at least 10$^4\,\rm{cm}^{-3}$, based on ionised-carbon line-ratios. If GLASS-z13 has a similar density, the non-detection of \oiii\ emission is to be expected, regardless of the metallicity of the ISM. A future approach to break the degeneracy between density and metallicity is to either obtain a density tracer from ionisation lines (following e.g., \citealt{Jiang2021}), or by constraining the ISM metallicity and density through a detection or upper limit on \ofivetwo\ emission (following for example the approach outlined in \citealt{Yang2021}).

\citet{Binggeli2021} recently presented an empirical relation between the \oiii-to-UV luminosity ratio and gas-phase metallicity of galaxies at $z>6$. Adopting a three $\sigma$ upper limit on the \oiii\ emission from GLASS-z13 of $1\times10^8\,\rm{L}_\odot$ and taking a UV magnitude of -21.0 \citep{Naidu2022}, I find an upper limit on the \oiii-to-UV luminosity ratio of $10^{-2.73}$. Following \citet{Binggeli2021} (their Figure 6), this is consistent with an upper limit on the oxygen abundance of approximately one-third solar (this conclusions remains qualitatively the same when adopting a line-width of 400 rather than 100 km s${^-1}$). These results agree with the hypothesis of a low-metallicity ISM based on the comparison of the \oiii\ luminosity--SFR ratio to CLOUDY modelling predictions.

Future JWST/NIRSpec spectroscopy will be able to confirm or reject the tentative high redshift of GLASS-z13. If GLASS-z13 indeed has a redshift of $z\approx 12-13$, the upper-limit on its \oiii\ line-emission may imply that GLASS-z13 has a low metallicity and/or dense ISM. This shows that even an ALMA \oiii\ non-detection may provide unique insights into the ISM properties and star-formation history of the first galaxies that formed in our Universe. 

\subsection{GLASS-z13 is a $z=4$--5 DSFG or quenched galaxy}
The second scenario to explain the non-detection of \oiii\ emission is that GLASS-z13 is a $z\approx3$ quenched or $z\approx4-6$ DSFG. The drop in the JWST/NIRCam photometry between F150W and F200W (see \citealt{Castellano2022} and \citealt{Naidu2022}) is then either driven by strong dust obscuration of UV emission from a $z\approx4-6$ DSFG,  or by the 4000 \AA\ break of a $z\sim3$ quenched galaxy. 

\citet{Zavala2022} recently presented 1.1 mm Northern Extended Millimeter Array (NOEMA) observations of a $z>17$ JWST galaxy candidate. They demonstrated that with the inclusion of the 1.1~mm continuum emission, the SED fit of this $z\sim17$ candidate  is more consistent with a $z\sim5$ DSFG than a $z>17$ solution. The stringent 3-$\sigma$ upper limit on the 1.2~mm continuum emission (which at $z\approx4-6$ would be close to the peak of the far-infrared SED) suggests that GLASS-z13 is unlikely a DSFG. Indeed, the 3-$\sigma$ upper limit at 1.2~mm corresponds to a total IR luminosity of $1.52\times10^{10}\,\rm{L}_\odot$ for a galaxy at $z=4$ (assuming a dust temperature of 35 K) and an upper limit on the SFR of only 2.25 $\rm{M}_\odot\,\rm{yr}^{-1}$ (following \citealt{Kennicutt2012}).

\citet{Naidu2022} explored the possibility of GLASS-z13 being a low-redshift galaxy by forcing a $z<6$ best-fit SED solution using the \texttt{EaZy} SED-fitting code \citep{Brammer2008}. They found that the JWST/NIRCam photometry can best be described by a $z\approx3$ quenched galaxy with a stellar mass of $10^{8-9}\,\rm{M}_\odot$ when adopting the $z<6$ constraint.  This solution is however unlikely, as based on this solution one would expect a $>5\sigma$ detection in the F150W band where no flux is present (see Figure~1 in \citealt{Naidu2022}). Future JWST/NIRSpec spectroscopy may be able to robustly confirm or rule out a $z\approx 3-6$ galaxy solution, while also accounting for the uncertainty in the zero-point of the NIRCam flux-calibration.

\subsection{GLASS-z13 has a high redshift outside of the frequency range covered}
A third explanation for the lack of detected \oiii\ emission is that GLASS-z13 is actually at a redshift $z>11$, but outside of the frequency range covered by the ALMA data presented. SED fitting of the JWST/NIRCam photometry with \texttt{Prospector} \citep{Leja2017, Leja2019, Johnson2021} by \citet{Naidu2022} shows that a redshift between $z=13.5$ and $z=15$ is also a possible solution. However, this solution is in  disagreement with the results from the \texttt{EaZy} \citep{Brammer2008} and \texttt{zphot} \citep{Fontana2000} fitting codes (see \citealt{Naidu2022} and \citealt{Castellano2022}). 

\section{Summary}
\label{sec:summary}
In this paper I presented publicly available ALMA band 6 DDT observations (2021.A.00020.S, PI: T. Bakx) of GLASS-z13, a $z\approx12-13$ galaxy candidate based on SED fitting of JWST/NIRCam photometry from 1 to 5 $\mu$m. The observations were designed to acquire a spectroscopic redshift for GLASS-z13, by searching for  \oiii\ emission in a contiguous frequency range corresponding to \oiii\ emission at redshifts 11.9 to 13.5. The main results include:
\begin{itemize}
\item No \oiii\ emission line associated to GLASS-z13 (nor in its vicinity) is detected in the frequency range corresponding to redshifts 11.9 to 13.5, nor is continuum emission at 1.2~mm detected.
\item The lack of detected \oiii\ emission associated to GLASS-z13 can be explained by three scenarios:
\begin{enumerate}
    \item  GLASS-z13 is at $z\approx12-13$, but has an \oiii\ luminosity fainter than the noise limit of the data presented in this work.
    \item GLASS-z13 is a $z\approx 3$ low-mass quenched galaxy, although the SED expected for such a galaxy is in disagreement with the JWST/NIRCam photometry \citep{Naidu2022}. Alternatively, it may be a $z\approx 4-6$ DSFG, although the stringent 1.2~mm continuum upper limit appears to rule out this scenario as well.
    \item  GLASS-z13 has a redshift in the range $z=13.5-15$, in agreement with the redshift solution of the \texttt{Prospector} SED fitting code, but ruled out by the \texttt{zphot} and \texttt{EaZy} codes \citep[see,][]{Naidu2022}.
\end{enumerate} 
\item If GLASS-z13 is at $z\approx12-13$, the \oiii\ upper-limit of $1\times10^8\,\rm{L}_\odot$ can be explained by a  low metallicity ($\sim 0.2\,\rm{Z}_\odot$) and/or dense ($\>100 \,\rm{cm}^{-3}$ or higher) ISM. A low metallicity is consistent with recent model predictions for the metallicity of galaxies at $z=12-13$ \citep{Wilkins2017,Ucci2021} and an empirical relation between the \oiii-to-UV luminosity ratio and oxygen abundance of \citet{Binggeli2021}.
\end{itemize}

The non-detection of the \oiii\ emission-line  makes it currently impossible to confirm or reject the $z\approx12-13$ photometric redshift of GLASS-z13. JWST/NIRSpec spectroscopy will be necessary for definite confirmation or rejection. If GLASS-z13 is at $z\approx12-13$ the presented ALMA observations provide new insights into the ISM properties of galaxies at $z\approx12-13$. This demonstrates the tremendous synergy between JWST and ALMA to study the existence and properties of the first galaxies to have formed in our Universe.

\begin{acknowledgements}
I thank Tony Mroczkowski and Michele Ginolfi for useful discussions and Tom Bakx for noticing an error in an earlier version of the paper. I thank the referee, Akio Inoue, for a constructive report. This paper makes use of the following ALMA data: ADS/JAO.ALMA\#2021.A.00020.S. ALMA is a partnership of ESO (representing its member states), NSF (USA) and NINS (Japan), together with NRC (Canada), MOST and ASIAA (Taiwan), and KASI (Republic of Korea), in cooperation with the Republic of Chile. The Joint ALMA Observatory is operated by ESO, AUI/NRAO and NAOJ. I made use of the following software packages: \texttt{matplotlib} \citep{Hunter2007}, \texttt{numpy} \citep{numpy}, \texttt{scipy} \citep{Virtanen2020}, \texttt{jupyter} \citep{Kluyver2016}, \texttt{Astropy} \citep{astropy}, \texttt{interferopy} \citep{Boogaard2021} and \texttt{CARTA} \citep{carta}. This paper made use of calibrated measurement sets provided by the European ALMA Regional Centre network \citep{Hatziminaoglou2015} through the calMS service \citep{Petry2020}.
\end{acknowledgements}   

% section summary_of_implications (end)

\bibliographystyle{aa}
\bibliography{mybib}

\appendix
\section{uv-tapered images}
\label{sec:appendix}
\begin{figure*}
	\centering
	    \includegraphics[width=\linewidth]{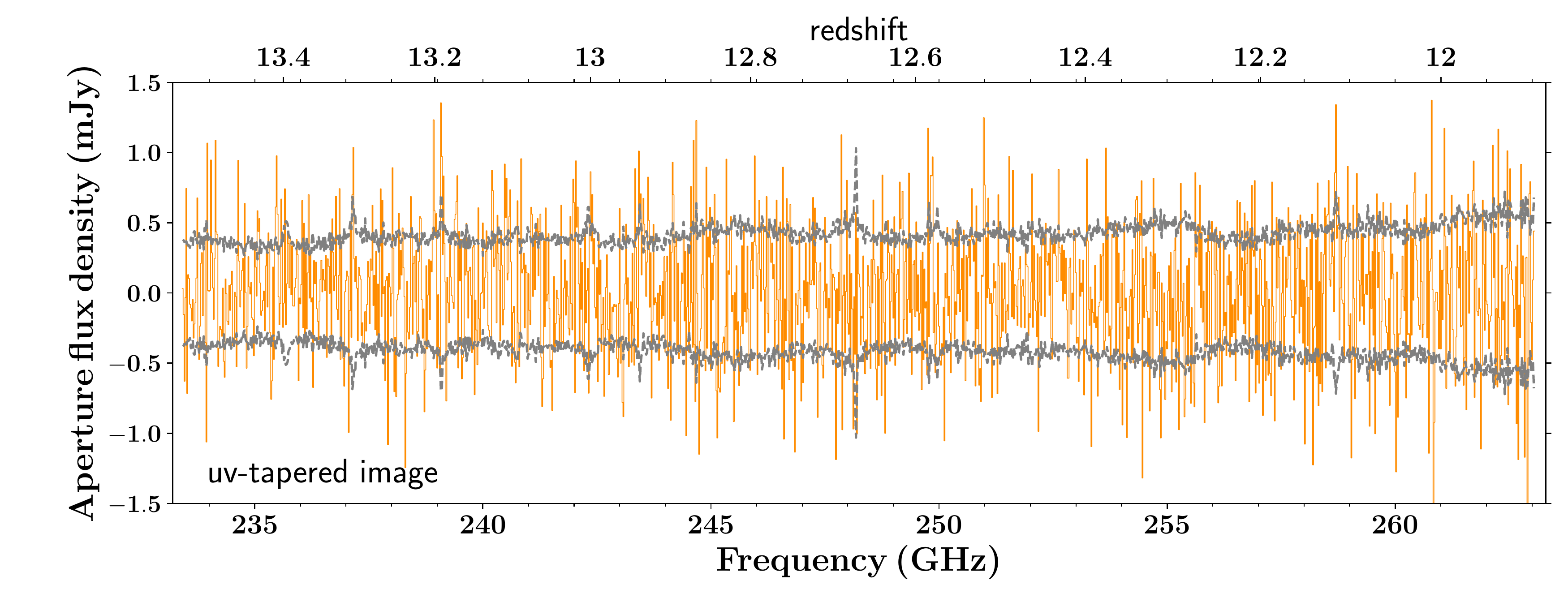}
     \caption{\label{fig:spectral_scan_tapered}The ALMA spectrum of GLASS-z13, extracted from the uv-tapered cube (with a taper of 0.6 arcsec). The spectrum was created by extracting the flux density within a circular aperture with a diameter of 1 arcsec around GLASS-z13. The orange bars indicate the spectrum, whereas the grey dashed lines correspond to the one-$\sigma$ noise limit of the data. No emission line is detected within the spectrum.}
\end{figure*}

In Figure~\ref{fig:spectral_scan_tapered} I show the integrated flux density spectrum of GLASS-z13 within an aperture with a diameter of 1 arcsec extracted from the uv-tapered cube. An aperture of 1 arcsec is slightly larger than the beam size of 0.8 arcsec of the uv-tapered image. No emission line is visible in this spectrum. Automated line finding using \texttt{Findclump} does not reveal any reliabl emission line within 10 arcsec of the location of GLASS-z13 either. These results are consistent with the results discussed in the main body of this work.

\begin{figure}
	    \includegraphics[width=\linewidth]{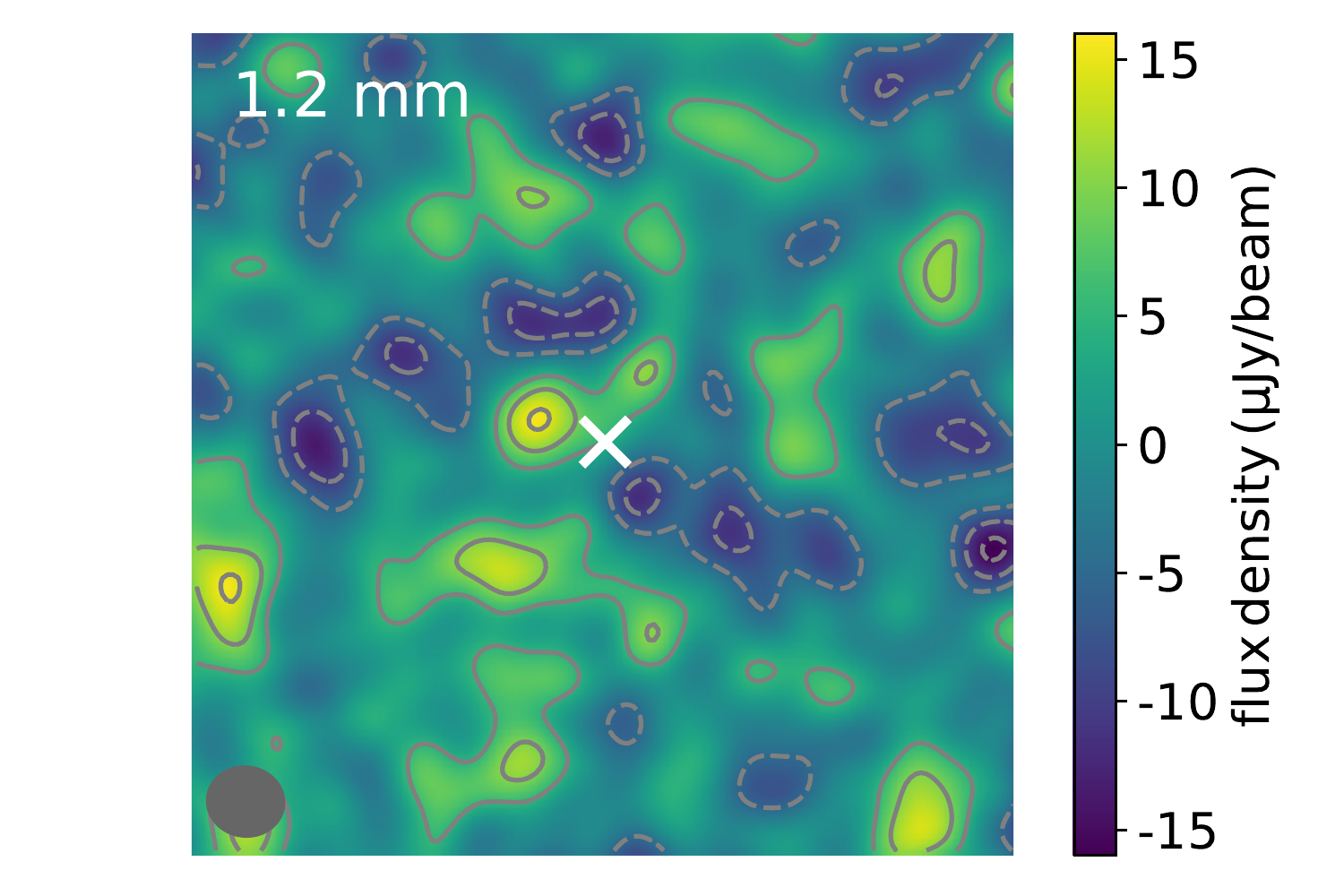}
     \caption{\label{fig:poststampTapered}An 8" $\times$ 8" cutout of the ALMA uv-tapered 1.2~mm continuum map around GLASS-z13 (with a taper of 0.6 arcsec). Contours indicate the -3, -2, -1, 1, 2 and 3 $\sigma$ levels. The beam of the ALMA 1.2~mm image is indicated with the grey ellipse in the bottom left corner, whereas the location of GLASS-z13 is indicated with a white cross. No continuum emission is detected at the location of GLASS-z13 in the uv-tapered image. }
\end{figure}

I present the the uv-tapered 1.2~mm continuum map around GLASS-z13 in Figure \ref{fig:poststampTapered}. No 1.2 mm continuum emission is visible in this image, in agreement with the results discussed in the main body of this work. 

\end{document}